\documentclass[aps,prl,reprint,showpacs,superscriptaddress]{revtex4-1}

\usepackage{graphicx,color,rotating,pifont}
\usepackage{amsmath,amssymb,bm}
\usepackage{ae}
\usepackage{pstricks}
\usepackage{dcolumn}
\usepackage{times}

\DeclareMathSymbol{\NS}{\mathord}{AMSb}{"4E}

\newcommand{\ket}[1]{\ensuremath{\,|{#1}\rangle}}

\newcommand{\matrixe}[3]{\ensuremath{\langle{#1}|\,{#2}\,|{#3}\rangle}}

\newcommand{\dmatrixe}[2]{\matrixe{#1}{#2}{#1}}

\newcommand{\comm}[2]{\ensuremath{[{#1},{#2}]}}

\newcommand{\op}[1]{\ensuremath{#1}}
\newcommand{\adj}[1]{\ensuremath{{{#1}}^{\dag}}}

\newcommand{\aO}{\ensuremath{\op{a}}}

\newcommand{\etaO}{\ensuremath{\op{\eta}}}

\newcommand{\aaO}{\ensuremath{\adj{\op{a}}}}

\newcommand{\AO}{\ensuremath{\op{A}}}

\newcommand{\HO}{\ensuremath{\op{H}}}

\newcommand{\PO}{\ensuremath{\op{P}}}

\newcommand{\totd}[2]{\ensuremath{ \frac{d {#1}} {d {#2}} }}
\newcommand{\nn}{\ensuremath{\bar{n}}}

\newcommand{\eMax}{\ensuremath{e_{\text{max}}}}

\newcommand{\EMax}{\ensuremath{E_{3\text{max}}}}
\newcommand{\NMax}{\ensuremath{N_{\text{max}}}}

\newcommand{\nuc}[2]{\ensuremath{{}^{#2}\mathrm{#1}}}

\newcommand{\fm}{\ensuremath{\,\text{fm}}}

\newcommand{\keV}{\ensuremath{\,\text{keV}}}
\newcommand{\MeV}{\ensuremath{\,\text{MeV}}}

\newcommand{\lamSRG}{\ensuremath{\lambda_\text{SRG}}}
\newcommand{\nord}[1]{\ensuremath{:\!#1:}}

\definecolor{FGViolet}{rgb}{0.61,0.32,0.61}
\definecolor{FGDarkBlue}{rgb}{0,0,0.6}
\definecolor{FGBlue}{rgb}{0,0,0.8}
\definecolor{FGLightBlue}{rgb}{0.2, 0.6, 0.8}
\definecolor{FGGreen}{rgb}{0.2,0.7,0.2}
\definecolor{FGLightGreen}{rgb}{0.4,1,0.4}
\definecolor{FGYellow}{rgb}{1,0.95,0}
\definecolor{FGOrange}{rgb}{0.95,0.5,0.1}
\definecolor{FGRed}{rgb}{0.8,0,0}
\definecolor{FGWhite}{rgb}{1,1,1}
\definecolor{FGLightGray}{rgb}{0.8,0.8,0.8}
\definecolor{FGGray}{rgb}{0.5,0.5,0.5}
\definecolor{FGDarkGray}{rgb}{0.3,0.3,0.3}
\definecolor{FGBlack}{rgb}{0,0,0}

\begin{document}
\title{\emph{Ab Initio} Calculations of Even Oxygen Isotopes with Chiral Two- Plus Three-Nucleon Interactions}

\author{H. Hergert}
\affiliation{The Ohio State University, Columbus, OH 43210, USA}
\email{Corresponding author. Electronic address: hergert.3@osu.edu}

\author{S. Binder}
\author{A. Calci}
\author{J. Langhammer}
\author{R. Roth}
\affiliation{Institut f\"ur Kernphysik, Technische Universit\"at Darmstadt,
D-64289 Darmstadt, Germany}

\date{\today}

\begin{abstract}
We formulate the In-Medium Similarity Renormalization Group (IM-SRG) for open-shell nuclei using a multi-reference formalism based on a generalized Wick theorem introduced in quantum chemistry. The resulting multi-reference IM-SRG (MR-IM-SRG) is used to perform the first \emph{ab initio} study of all even oxygen isotopes with chiral NN and 3N Hamiltonians, from the proton to the neutron drip lines. We obtain an excellent reproduction of experimental ground-state energies with quantified uncertainties, which is validated by results from the Importance-Truncated No-Core Shell Model and the Coupled Cluster method. The agreement between conceptually different many-body approaches and experiment highlights the predictive power of current chiral two- and three-nucleon interactions, and establishes the MR-IM-SRG as a promising new tool for \emph{ab initio} calculations of medium-mass nuclei far from shell closures.
\end{abstract}


\maketitle

%%%%%%%%%%%%%%%%%%%%%%%%%%%%%%%%%%%%%%%%%%%%%%%%%%%%%%%%%%%%%%%%%%%%%%
%%%%%%%%%%%%%%%%%%%%%%%%%%%%%%%%%%%%%%%%%%%%%%%%%%%%%%%%%%%%%%%%%%%%%%
%%%%%%%%%%%%%%%%%%%%%%%%%%%%%%%%%%%%%%%%%%%%%%%%%%%%%%%%%%%%%%%%%%%%%%
\clearpage

%%%%%%%%%%%%%%%%%%%%%%%%%%%%%%%%%%%%%%%%%%%%%%%%%%%%%%%%%%%%%%%%%%%%%%
%%%%%%%%%%%%%%%%%%%%%%%%%%%%%%%%%%%%%%%%%%%%%%%%%%%%%%%%%%%%%%%%%%%%%%
%%%%%%%%%%%%%%%%%%%%%%%%%%%%%%%%%%%%%%%%%%%%%%%%%%%%%%%%%%%%%%%%%%%%%%
\paragraph{Introduction.}
Neutron-rich nuclei are the focus of the experimental program of current and next-generation rare isotope facilities. Emerging phenomena such as halos or neutron skins make these nuclei ideal laboratories to study nuclear interactions in delicately tuned scenarios, and motivate the use of \emph{ab initio} many-body calculations to provide their description from first principles. Such calculations make it possible to confront modern nuclear Hamiltonians from chiral effective field theory (EFT) \cite{Epelbaum:2009ve,Machleidt:2011bh} with a wealth of data beyond few-body systems. 

For light nuclei, the \emph{ab initio} No-Core Shell Model (NCSM) \cite{Navratil:2000nm,Barrett:2013oq} provides the capabilities for studies of isotopic chains, but for medium-mass nuclei this approach is not feasible because of its large computational effort. Many-body techniques with more modest computational scaling, such as the Coupled Cluster (CC) \cite{Jansen:2011tg,Hagen:2012oq,Hagen:2012nx} or Self-Consistent Green's Function methods \cite{Barbieri:2009kx,Cipollone:2013uq}, can be used to probe nuclei in the vicinity of shell closures, but are not applicable for open-shell nuclei far from shell closures. For such nuclei, a self-consistent Gor'kov formalism was developed recently \cite{Soma:2011vn,Soma:2013ys}, but this approach is currently limited to second-order terms in the many-body perturbation expansion.
  
In this Letter, we describe the extension of the In-Medium Similarity Renormalization Group (IM-SRG) framework of Refs.~\cite{Tsukiyama:2011uq,Hergert:2012nb} to open-shell nuclei by means of a multi-reference formulation. We use the resulting MR-IM-SRG and two other many-body approaches, the Importance-Truncated No-Core Shell Model (IT-NCSM) and the CC method, to perform the first \emph{ab initio} study of all even oxygen isotopes with chiral NN+3N Hamiltonians.

%%%%%%%%%%%%%%%%%%%%%%%%%%%%%%%%%%%%%%%%%%%%%%%%%%%%%%%%%%%%%%%%%%%%%%
%%%%%%%%%%%%%%%%%%%%%%%%%%%%%%%%%%%%%%%%%%%%%%%%%%%%%%%%%%%%%%%%%%%%%%
% SECTION %%%%%%%%%%%%%%%%%%%%%%%%%%%%%%%%%%%%%%%%%%%%%%%%%%%%%%%%%%%%
%%%%%%%%%%%%%%%%%%%%%%%%%%%%%%%%%%%%%%%%%%%%%%%%%%%%%%%%%%%%%%%%%%%%%%
\paragraph{Formalism.}
The main tools for the derivation of the MR-IM-SRG are the generalized normal-ordering and Wick theorem by Kutzelnigg and Mukherjee \cite{Kutzelnigg:1997fk}. We write a string of creation and annihilation operators in tensorial form,
\begin{equation}
  \AO^{1\ldots k}_{l\ldots N}\equiv\aaO_1\ldots\aaO_k\aO_N\ldots\aO_l\,,
\end{equation}
and expand it in terms of components that are normal-ordered with respect to an arbitrary reference state $\ket{\Phi}$ \cite{Kutzelnigg:1997fk,Kong:2010kx,Shavitt:2009}. We obtain
\begin{align}\label{eq:nord}
  \AO^{1\ldots k}_{l\ldots N} 
    &=\,\nord{\AO^{1\ldots k}_{l\ldots N}} + \lambda^{1}_l\nord{\AO^{23\ldots k}_{mn\ldots N}} - \lambda^{1}_{m}\nord{\AO^{23\ldots k}_{ln\ldots N}} + \ldots \notag\\
    &\hphantom{=} + (\lambda^{1}_{l}\lambda^{2}_{m} - \lambda^{1}_{m}\lambda^{2}_{l} + \lambda^{12}_{lm})\nord{\AO^{3\ldots k}_{n\ldots N}} + \ldots\,,
\end{align}
where $:.:$ indicates normal-ordering, and we have introduced irreducible one- and two-body density matrices $\lambda^{(1)}$ and $\lambda^{(2)}$:
\begin{align}
  \lambda^{1}_{2}   \equiv \dmatrixe{\Phi}{\!\AO^{1}_{2}}\,, \quad\!
  \lambda^{12}_{34} \equiv \dmatrixe{\Phi}{\!\AO^{12}_{34}} - \lambda^{1}_{2}\lambda^{3}_{4} 
                       + \lambda^{1}_{3}\lambda^{2}_{4}\,.
\end{align}
The particle rank of the irreducible density matrices is evident from the single-particle indices. Generally, up to $n$-body irreducible density matrices $\lambda^{(n)}$ appear in the expansion of an $n$-body operator, which are defined recursively in terms of density matrices of lower rank and encode information about $n$-body correlations in the reference state \cite{Kutzelnigg:1997fk}. For an independent-particle state, all matrices except $\lambda^{(1)}$ vanish. 

Products of normal-ordered operators can be expanded by means of a generalized Wick theorem (GWT), e.g.,
\begin{align} 
  % &\nord{\AO^{12}_{56}}\nord{\AO^{34}_{78}} \notag\\
  % &= \;\nord{\AO^{1234}_{5678}} + \lambda^{1}_{7}\nord{\AO^{234}_{568}} - \xi^{3}_{5}\nord{\AO^{124}_{678}} + \ldots \notag\\
  % &\hphantom{=}+\left(\lambda^{1}_{7}\lambda^{2}_{8} - \lambda^{1}_{8}\lambda^{2}_{7} + \lambda^{12}_{78}\right) \nord{\AO^{34}_{56}} 
  % - \lambda^{12}_{57}\nord{\AO^{34}_{68}} + \ldots\,,
  &\nord{\AO^{12}_{56}}\nord{\AO^{34}_{78}} 
  \;= \;\nord{\AO^{1234}_{5678}} + \lambda^{1}_{7}\nord{\AO^{234}_{568}} - \xi^{3}_{5}\nord{\AO^{124}_{678}} + \ldots \notag\\
  &\hphantom{=}+\left(\lambda^{1}_{7}\lambda^{2}_{8} - \lambda^{1}_{8}\lambda^{2}_{7} + \lambda^{12}_{78}\right) \nord{\AO^{34}_{56}} 
  - \lambda^{12}_{57}\nord{\AO^{34}_{68}} + \ldots\,,
\end{align}
where $\xi^{1}_{2} \equiv \lambda^{1}_{2} - \delta^{1}_{2}$ \cite{Kong:2010kx}. In addition to simple contractions containing $\lambda^{(1)}$ and $\xi^{(1)}$ which also occur in the standard Wick theorem, we obtain terms involving $\lambda^{(2)},\ldots, \lambda^{(n)}$. Each density matrix must have at least one index from each of the operators in the product --- other terms vanish due to the initial normal-ordering \eqref{eq:nord} \cite{Kutzelnigg:1997fk}. In the following, we work in natural orbitals, i.e., the eigenbasis of $\lambda^{(1)}$, where
\begin{equation}
  \lambda^{1}_{2}=n_{1}\delta^{1}_{2}\,,\quad\xi^{1}_{2}=-\bar{n}_{1}\delta^{1}_{2}\equiv-(1-n_{1})\delta^{1}_{2}\,,
\end{equation}
and the eigenvalues are the occupation numbers $0\leq n_a\leq 1$.

We now consider the IM-SRG operator flow equation
\begin{equation}\label{eq:flow}
  \totd{}{s}\HO(s) = \comm{\etaO(s)}{\HO(s)}\,.
\end{equation}
By integrating Eq.~\eqref{eq:flow}, we generate a continuous unitary transformation that decouples the ground-state of the Hamiltonian $\HO(s)$ from excitations, and solve the many-body problem \cite{Tsukiyama:2011uq, Hergert:2012nb}.
% which is used to solve the many-body problem by a continuous unitary transformation that decouples the ground-state of the Hamiltonian $\HO(s)$ from excitations \cite{Tsukiyama:2011uq, Hergert:2012nb}. 
Suppressing the flow parameter $s$ for brevity, we apply the generalized normal-ordering to $\HO$ and the generator $\etaO$, and evaluate the commutator using the GWT to obtain the MR-IM-SRG flow equations:
\begin{align}
  \totd{E}{s} &=     
  \sum_{ab}(n_{a}-n_{b})\left(\eta^{a}_{b}f^{b}_{a}-f^{a}_{b}\eta^{b}_{a}\right)\notag\\
  &\hphantom{=}
  +\frac{1}{4}\sum_{abcd}\left(\eta^{ab}_{cd}\Gamma^{cd}_{ab}-\Gamma^{ab}_{cd}\eta^{cd}_{ab}\right)n_{a}n_{b}\bar{n}_{c}\bar{n}_{d}
    \notag\\
  &\hphantom{=}
    +\frac{1}{4}\sum_{abcd}\left(\totd{}{s}\Gamma^{ab}_{cd}\right)\lambda^{ab}_{cd}\,,
  \label{eq:mr_flow_0b_tens}
\end{align}
\begin{align}
  \totd{}{s}f^{1}_{2} &=
    \sum_{a}\eta^{1}_{a}f^{a}_{2} +\sum_{ab}\eta^{a}_{b}\Gamma^{b1}_{a2}(n_{a}-n_{b})
  \notag\\
  &\hphantom{=}
  +\frac{1}{2}\sum_{abc}\eta^{1a}_{bc}\Gamma^{bc}_{2a}\left(n_{a}\bar{n}_{b}\bar{n}_{c}+\bar{n}_{a}n_{b}n_{c}\right)
  \notag\\
  &\hphantom{=}
      +\frac{1}{4}\sum_{abcde}\eta^{1a}_{bc}\Gamma^{de}_{2a}\lambda^{de}_{bc} 
    +\sum_{abcde}\eta^{1a}_{bc}\Gamma^{be}_{2d}\lambda^{ae}_{cd}
  \notag\\
  &\hphantom{=}
      -\frac{1}{2}\sum_{abcde}\left(\eta^{1a}_{2b}\Gamma^{cd}_{ae}\lambda^{cd}_{be}
                                    -\eta^{1a}_{2b}\Gamma^{bc}_{de}\lambda^{ac}_{de}\right)
    -\left[\eta \leftrightarrow f,\Gamma\right]\,,
  \label{eq:mr_flow_1b_tens}
\end{align}
\begin{align}
  \totd{}{s}\Gamma^{12}_{34}&=  
  \sum_{a}\left(\eta^{1}_{a}\Gamma^{a2}_{34}+\eta^{2}_{a}\Gamma^{1a}_{34}-\eta^{a}_{3}\Gamma^{12}_{a4}-\eta^{a}_{4}\Gamma^{12}_{3a}
  \right.\notag\\
  &\hphantom{=}\left.
  \qquad-f^{1}_{a}\eta^{a2}_{34}-f^{2}_{a}\eta^{1a}_{34}+f^{a}_{3}\eta^{12}_{a4}+f^{a}_{4}\eta^{12}_{3a}\right)
  \notag\\
  &\hphantom{=}
    +\frac{1}{2}\sum_{ab}\left(\eta^{12}_{ab}\Gamma^{ab}_{34}-\Gamma^{12}_{ab}\eta^{ab}_{34}\right)
     \left(1-n_{a}-n_{b}\right)
  \notag\\
  &\hphantom{=}
    +\sum_{ab}(n_{a}-n_{b})\left(\left(\eta^{1a}_{3b}\Gamma^{2b}_{4a}-\Gamma^{1a}_{3b}\eta^{2b}_{4a}\right)
     -\left[1\leftrightarrow2\right]\right)\,,
  \label{eq:mr_flow_2b_tens}
\end{align}
where $E=\matrixe{\Phi}{\HO}{\Phi}$, and the one- and two-body parts of $\HO$, denoted by $f$ and $\Gamma$, contain in-medium contributions from the 3N interaction because of the normal ordering \cite{Tsukiyama:2011uq,Hergert:2012nb}. The symbol $[\eta\leftrightarrow f,\Gamma]$ in Eq.~\eqref{eq:mr_flow_1b_tens} indicates an interchange of the one- and two-body parts of $\eta$ and $\HO$. To close the system of flow equations \eqref{eq:mr_flow_0b_tens}--\eqref{eq:mr_flow_2b_tens}, we truncate three-body operators \cite{Hergert:2012nb} and a term containing $\lambda^{(3)}$ in the energy flow equation \eqref{eq:mr_flow_0b_tens}. We refer to this truncation as MR-IM-SRG(2). 

By integrating Eqs.~\eqref{eq:mr_flow_0b_tens}--\eqref{eq:mr_flow_2b_tens}, we perform a non-perturbative resummation of the Many-Body Perturbation series \cite{Tsukiyama:2011uq, Hergert:2012nb}. The flowing two-body vertex is RG-improved by Eq.~\eqref{eq:mr_flow_2b_tens}, e.g., with contributions from generalized ladder (3rd line) and ring diagrams (4th line), which in turn generate corrections to the ground-state energy when $\Gamma$ is inserted in Eq.~\eqref{eq:mr_flow_0b_tens} \cite{Hergert:2012nb}.

As our default choice for the generator, we use the ansatz of White \cite{White:2002fk,Hergert:2012nb}. The required matrix elements of the Hamiltonian, such as $\matrixe{\Phi}{\HO\nord{\AO^{12}_{34}}}{\Phi}$, which couple the reference state to excitations, or $\matrixe{\Phi}{\nord{\AO^{34}_{12}}\HO\nord{\AO^{12}_{34}}}{\Phi}$, which enter the energy denominators, can be evaluated using the generalized normal ordering. This yields
\begin{align}
  \eta^{1}_{2} &= \frac{\nn_1 n_2 f^1_2}{\nn_1f^{1}_{1} -n_2f^{2}_{2} + \nn_1 n_2\Gamma^{12}_{12}} - \left[1 \leftrightarrow 2 \right] + \ldots\,,\label{eq:eta1b}\\
  \eta^{12}_{34} &= \frac{\nn_1\nn_2 n_3n_4 \Gamma^{12}_{34}}{\nn_1f^{1}_{1}+\nn_2f^{2}_{2} - n_3f^{3}_{3} - n_4f^{4}_{4} + G^{12}_{34}} - \left[(12) \leftrightarrow (34) \right] \notag\\ &\hphantom{=}+ \ldots\,,\label{eq:eta2b}
\end{align}
where 
\begin{align}
  G^{12}_{34} &= \nn_1\nn_2\Gamma^{12}_{12} + n_3n_4\Gamma^{34}_{34} \notag\\
        &\hphantom{=}- \left(\nn_1n_3\Gamma^{13}_{13} + \nn_2n_4\Gamma^{24}_{24} + \left[1\leftrightarrow2\right]\right)\,.
\end{align}
The dots in Eqs.~\eqref{eq:eta1b} and \eqref{eq:eta2b} indicate terms that are linear in $\lambda^{(2)}$. Terms containing $\lambda^{(n\geq3)}$ or nonlinear powers of $\lambda^{(2)}$ are truncated.

In cases where the flow stalls due to small energy denominators, we use Wegner's generator $\etaO=\comm{\HO}{\HO^{od}}$ as a fall-back, defining the one- and two-body parts of the off-diagonal Hamiltonian $\HO^{od}$ as
\begin{align}
  (f^{od})^{1}_{2} &= \nn_1 n_2 f^1_2 + \left[1 \leftrightarrow 2 \right]\,,\notag\\
  (\Gamma^{od})^{12}_{34} &= \nn_1\nn_2 n_3n_4 \Gamma^{12}_{34} + \left[(12) \leftrightarrow (34) \right]\,.
\end{align}
This generator is free of numerical instabilities but less efficient because the flow equations become stiff \cite{Hergert:2012nb, Tsukiyama:2011uq}. In the limit of a single Slater determinant reference state, both generators reduce to the forms used for closed-shell nuclei in \cite{Hergert:2012nb, Tsukiyama:2011uq}.

We obtain a reference state for each nucleus by solving the Hartree-Fock-Bogoliubov (HFB) equations, and projecting the resulting state on proton and neutron number, $\ket{\Phi} = \PO_N\PO_Z \ket{\text{HFB}}$ \cite{Ring:1980bb}. This choice allows us to enforce spherical symmetry in calculations for even nuclei \cite{Perez-Martin:2008zf}, and greatly increases the single-particle basis sizes we can treat. The natural-orbital basis of $\ket{\Phi}$ is the usual canonical basis of the HFB vacuum, allowing us to use analytic expressions for the density matrices \cite{Sheikh:2000xx}.

The MR-IM-SRG method can be extended systematically by improving the truncation scheme: One would include $3,\ldots,A$-body operators when Eq.~\eqref{eq:flow} is expanded in normal-ordered components, as well as additional terms involving irreducible density matrices. While the number of flow equations is the same as in the single-reference case, their complexity grows much more rapidly due to additional terms from the generalized normal ordering \cite{Kutzelnigg:1997fk,Tsukiyama:2011uq,Hergert:2012nb}.

%%%%%%%%%%%%%%%%%%%%%%%%%%%%%%%%%%%%%%%%%%%%%%%%%%%%%%%%%%%%%%%%%%%%%%
% SECTION %%%%%%%%%%%%%%%%%%%%%%%%%%%%%%%%%%%%%%%%%%%%%%%%%%%%%%%%%%%%
%%%%%%%%%%%%%%%%%%%%%%%%%%%%%%%%%%%%%%%%%%%%%%%%%%%%%%%%%%%%%%%%%%%%%%
\paragraph{Calculation Details.}
Reference states for the MR-IM-SRG calculation are obtained by solving the HFB equations in 15 major harmonic-oscillator (HO) shells, and projecting the resulting state on good proton and neutron numbers \cite{Hergert:2009zn,Hergert:2012nb}. For the 3N interaction, the sum of the HO energy quantum numbers of a 3N basis state is limited by $e_1+e_2+e_3\leq\EMax=14$, as discussed in \cite{Binder:2013zr,Hergert:2012nb}. Reducing $\EMax$ from 14 to 12 changes the MR-IM-SRG(2) ground-state energies for oxygen isotopes by less than 1\% for the Hamiltonians used in this work. The intrinsic NN+3N Hamiltonian is normal-ordered with respect to the reference state, and the residual normal-ordered 3N interaction term is discarded, leading to the normal-ordered two-body approximation (NO2B), which is found to overestimate oxygen binding energies by about 1\% \cite{Hergert:2012nb,Binder:2013zr}.

In this Letter, we use the same nuclear Hamiltonians as in our recent IM-SRG and CC studies \cite{Roth:2012qf,Hergert:2012nb,Binder:2013zr}: The NN interaction is the chiral N$^3$LO interaction by Entem and Machleidt, with cutoff $\Lambda_\text{NN}=500\,\MeV/c$ \cite{Entem:2002sd,Machleidt:2011bh}. Our standard three-body Hamiltonian is a local N$^2$LO 3N interaction with initial cutoff $\Lambda_\text{3N}=400\,\MeV/c$. The resolution scale of the Hamiltonian is lowered to $\lamSRG=1.88,\ldots,2.24\,\fm^{-1}$ by means of an SRG evolution in three-body space \cite{Jurgenson:2009bs,Roth:2011kx,Roth:2012vn}. Hamiltonians which only contain SRG-induced 3N forces are referred to as NN+3N-induced, those also containing an initial 3N interaction as NN+3N-full. 

\begin{figure}[t]
  % \setlength{\unitlength}{0.5\columnwidth}
  % \begin{picture}(2.0000,1.3500)
  %   \put(0.0400,0.0000){\input{fig/chi2b3b400_srg0625_O18}}
  %   \put(1.0000,0.0000){\input{fig/chi2b3b400_srg0625_O26}}
  % \end{picture}
  % \vspace{-20pt}
  \includegraphics[width=0.95\columnwidth]{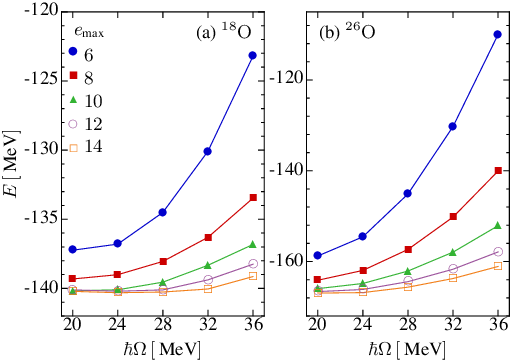}
  \vspace{-10pt}
  \caption{\label{fig:convergence}(Color online) Convergence of the MR-IM-SRG(2) ground-state energies of $\nuc{O}{18}$ and $\nuc{O}{26}$ with respect to the single-particle basis size $\eMax$, for the NN+3N-full Hamiltonian at $\lamSRG=2.0\,\fm^{-1}$.}
\end{figure}

In Fig.~\ref{fig:convergence}, we illustrate the convergence of the MR-IM-SRG(2) ground-state energies for $\nuc{O}{18}$ and $\nuc{O}{26}$ with respect to the single-particle basis size. At the optimal $\hbar\Omega$, the change in the ground-state energy is 0.1\% when we increase the basis from $\eMax=12$ to 14. This rapid convergence is representative for all Hamiltonians used in this work.

%%%%%%%%%%%%%%%%%%%%%%%%%%%%%%%%%%%%%%%%%%%%%%%%%%%%%%%%%%%%%%%%%%%%%%
% SECTION %%%%%%%%%%%%%%%%%%%%%%%%%%%%%%%%%%%%%%%%%%%%%%%%%%%%%%%%%%%%
%%%%%%%%%%%%%%%%%%%%%%%%%%%%%%%%%%%%%%%%%%%%%%%%%%%%%%%%%%%%%%%%%%%%%%
\paragraph{Results.}
\begin{figure}[t]
  % \setlength{\unitlength}{\columnwidth}
  % \hspace{-15pt}
  % \begin{picture}(1.0000,0.8000)
  %   \put(0.0000,0.0000){\input{fig/chi2b3bXXX_srgXXXX_OXX_Eint}}
  %   \put(0.4700,0.3900){\setlength{\unitlength}{0.5\columnwidth}\input{fig/chi2b3bXXX_srgXXXX_OXX_Eint_inset}}
  % \end{picture}  
  % \vspace{-30pt}
  \includegraphics[width=0.95\columnwidth]{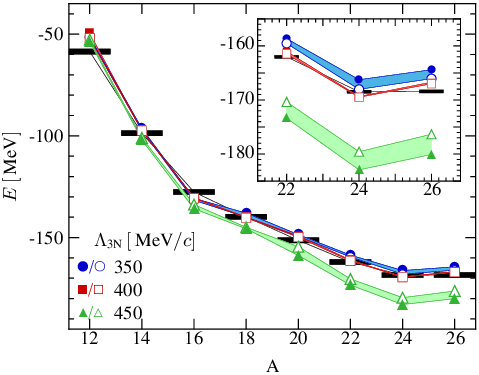}  
  \vspace{-5pt}
  \caption{\label{fig:lambda}(Color online) Dependence of the MR-IM-SRG(2) oxygen ground-state energies for the NN+3N-full Hamiltonian on the resolution scale and the initial cutoff $\Lambda_\text{3N}$. For each $\Lambda_\text{3N}$, the band is obtained by varying $\lamSRG$ from $2.24$ (open symbols) to $1.88\,\fm^{-1}$ (closed symbols). Experimental values are indicated by black bars \cite{Audi:2002af,Hoffman:2008ly}.}
\end{figure}

In Fig.~\ref{fig:lambda}, we show MR-IM-SRG(2) ground-state energies of the even oxygen isotopes for NN+3N-full Hamiltonians with initial cutoffs $\Lambda_\text{3N}=350,400$ and $450\,\MeV/c$. For the 3N low-energy constants, we use a fixed $c_D=-0.2$, and $c_E=0.205,0.098$, and $-0.016$, respectively, which are fit to the $\nuc{He}{4}$ binding energy in NCSM calculations \cite{Roth:2012qf,Roth:2012vn}. For the NN+3N-full Hamiltonian with $\Lambda_\text{3N}=400\,\MeV/c$, we achieve an excellent reproduction of experimental data all the way to the neutron drip line at $\nuc{O}{24}$ \cite{Hoffman:2008ly}, with deviations of 1-2\%. Recent experiments place the $\nuc{O}{26}$ ground-state resonance at $E_x\lesssim150\,\keV$ above the $\nuc{O}{24}$ ground-state energy \cite{Lunderberg:2012cr,Caesar:2012uq}. We slightly overestimate this energy in our calculation because the HO basis expansion of our single-particle wave functions is ill-suited to the description of resonances and other continuum states. The inset in Fig.~\ref{fig:lambda} illustrates that the correct drip-line systematics is independent of $\lamSRG$ in the studied range and also robust against variations of the cutoff $\Lambda_\text{3N}$. This suggests that the long-range part of the two-pion exchange (2PE) 3N interaction, which remains unchanged as we lower $\Lambda_\text{3N}$, is key to obtaining the proper isotopic trends. The 2PE contribution has significant spin-orbit and tensor terms, and is therefore important for the evolution of the shell structure along the isotopic chain, as also demonstrated in other studies, e.g. \cite{Otsuka:2010cr}. 

Let us now discuss the effect of varying the resolution scale. As discussed in \cite{Hergert:2012nb,Binder:2013zr}, the $\lamSRG$-dependence of our energies is the net result of omitted induced 4N interactions, the $\EMax$ cut, and the MR-IM-SRG(2) truncation of the many-body expansion, while the effect of the NO2B approximation is found to be independent of $\lamSRG$. 

For $\Lambda_\text{3N}=350\,\MeV/c$ we do not expect significant induced 4N interactions \cite{Roth:2012vn}. As $\lamSRG$ is reduced, we capture additional repulsive 3N strength in matrix elements with $e_1+e_2+e_3\leq\EMax$. We also speed up the convergence of the many-body expansion and reduce the error due to the MR-IM-SRG(2) truncation, but for the resolution scales considered here, this effect is already saturated. In total, we find a slight artificial increase of the ground-state energies as we lower $\lamSRG$ \cite{Hergert:2012nb}. 

For our standard choice $\Lambda_\text{3N}=400\,\MeV/c$, effects from omitted 4N interactions, the $\EMax$ cut, and the many-body truncation cancel, and the $\lamSRG$-dependence of the energies in Fig.~\ref{fig:lambda} is extremely weak \cite{Hergert:2012nb}. The omission of 4N interactions becomes the dominant source of uncertainty as we increase $\Lambda_\text{3N}$ to $450\,\MeV/c$, resulting in an enhanced $\lamSRG$-dependence of the ground-state energies of the heavier oxygen isotopes. This is consistent with the even stronger $\lamSRG$-dependence for $\Lambda_\text{3N}=500\,\MeV/c$ observed in Refs.~\cite{Roth:2011kx,Roth:2012qf,Roth:2012vn}. 

\begin{figure}[t]
  \includegraphics[width=0.98\columnwidth]{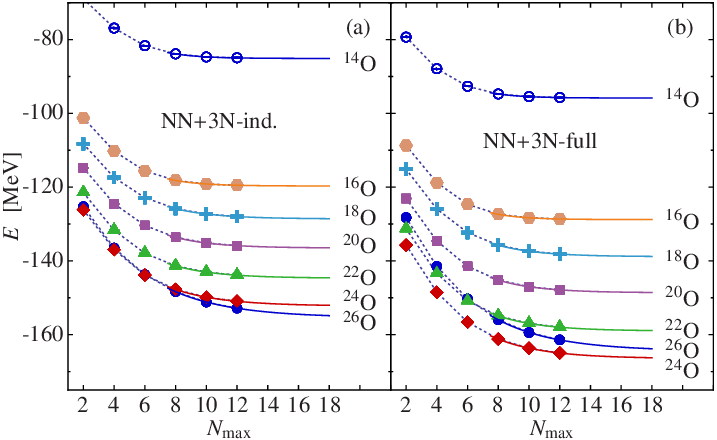}
  \vspace{-5pt}
  \caption{\label{fig:extrapol}(Color online) IT-NCSM ground-state energies of the even oxygen isotopes for the NN+3N-induced (a) and NN+3N-full Hamiltonians (b) at $\lamSRG=1.88\,\fm^{-1}$. Solid lines indicate the energy extrapolation based on $\NMax=8-12$ data, dotted lines guide the eye for smaller $\NMax$. Uncertainties due to the importance truncation are smaller than the symbols used to represent the data. All energies are obtained at optimal  $\hbar\Omega$.}
\end{figure}

To assess the quality of our MR-IM-SRG(2) ground-state energies, we compare them to results from the IT-NCSM, which yields the exact NCSM results within quantified uncertainties from the importance truncation \cite{Roth:2009eu,Roth:2011kx}. In the IT-NCSM calculations, we use the full 3N interaction without NO2B approximation, and the $\EMax$ cut is naturally compatible with the IT-NCSM model space truncation \cite{Hergert:2012nb}. In Fig.~\ref{fig:extrapol} we show the convergence of the oxygen ground-state energies for the NN+3N-induced and NN+3N-full Hamiltonians as a function of $\NMax$, along with exponential fits which extrapolate $\NMax\to\infty$ \cite{Roth:2007fk,Roth:2009eu,Roth:2011kx}. With the exception of $\nuc{O}{26}$, all isotopes converge well, and the uncertainties of the threshold and model spaces truncations of the IT-NCSM results are typically about 1 MeV. For $\nuc{O}{26}$, the rate of convergence is significantly worse, which is expected due to the resonance nature of this ground state.

The neutron-rich oxygen isotopes are the heaviest nuclei studied so far in the IT-NCSM with full $3N$ interactions. For $\nuc{O}{26}$, the computation of the complete $\NMax$ sequence shown in Fig. \ref{fig:extrapol} requires about 200,000 CPU hours. In contrast, a corresponding sequence of single-particle basis sizes in the MR-IM-SRG requires only about 3,000 CPU hours on a comparable system. Overall, the method scales polynomially with $\mathcal{O}(N^6)$ to larger basis sizes $N$, which makes it ideally suited for the description of medium- and heavy-mass nuclei. 

\begin{figure}[t]
  % \setlength{\unitlength}{0.95\columnwidth}
  % \hspace{-25pt}
  % \begin{picture}(1.0000,1.2200)
  %   \put(0.0000,0.5100){\input{fig/chi2b3bi_srg0800_OXX_methods_new}}
  %   \put(0.0000,0.0000){\input{fig/chi2b3b400_srg0800_OXX_methods_new}}
  % \end{picture}
  % \vspace{-30pt}
  \includegraphics[width=0.9\columnwidth]{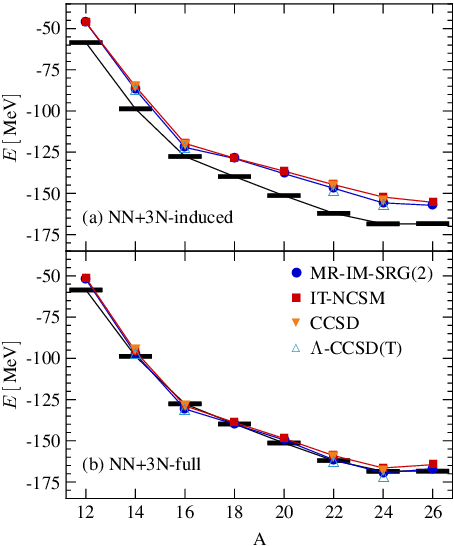}
  \vspace{-5pt}
  \caption{\label{fig:oxygen}(Color online) Oxygen ground-state energies for the NN+3N-induced (top) and NN+3N-full (bottom) Hamiltonian with $\Lambda_\text{3N}=400\,\MeV/c$. MR-IM-SRG(2), CCSD, and $\Lambda$-CCSD(T) results are obtained at optimal $\hbar\Omega$, using 15 major oscillator shells and $\EMax=14$. The IT-NCSM energies are extrapolated to infinite model space. Experimental values are indicated by black bars \cite{Audi:2002af,Hoffman:2008ly}.}
\end{figure}

In Fig.~\ref{fig:oxygen}, we compare the MR-IM-SRG(2) and IT-NCSM ground-state energies of the oxygen isotopes, for the NN+3N-induced and NN+3N-full Hamiltonians with $\lamSRG=1.88\,\fm^{-1}$ to experiment. For the latter, the overall agreement between the two very different many-body approaches and experiment is striking: Except for slightly larger deviations in $\nuc{O}{12}$ and $\nuc{O}{26}$, we reproduce experimental binding energies within 2-3$\,\MeV$. This is a remarkable demonstration of the predictive power of current chiral NN+3N Hamiltonians, at least for ground-state energies. For further confirmation, we perform CC calculations with singles and doubles (CCSD), as well as perturbative triples ($\Lambda$-CCSD(T)) \cite{Shavitt:2009,Taube:2008kx,Hagen:2010uq,Binder:2013zr} for oxygen isotopes with sub-shell closures. Using the same Hamiltonians in NO2B approximation, the MR-IM-SRG energies are bracketed by the CC results, and similar to the $\Lambda$-CCSD(T) values, consistent with the closed-shell results discussed in \cite{Hergert:2012nb}. 

For the NN+3N-induced calculation, which should be compared to calculations with the bare chiral NN interaction \cite{Hagen:2012oq}, the reproduction of experimental trends fails, and the neutron drip line is predicted at the wrong mass, because $\nuc{O}{26}$ is bound with respect to $\nuc{O}{24}$. This illustrates the crucial importance of the chiral 3N interaction for a proper description of the structure of neutron-rich nuclei \cite{Otsuka:2010cr}.

Let us now address the uncertainties of our results. The MR-IM-SRG(2) energies lie 1.5--2\% below the IT-NCSM results. About 1\% of this deviation is caused by the NO2B approximation. The uncertainty due to the $\EMax$ cut is less than 1\% at low $\lamSRG$. While these uncertainties exhaust the greater part of the 1.5--2\% deviation between MR-IM-SRG(2) and IT-NCSM, and suggest a very small uncertainty due to the many-body truncation, we assume a more conservative many-body truncation error of 1--1.5\%, and an overall uncertainty of our oxygen energies at the level of 3--3.5\%, consistent with our closed-shell IM-SRG calculations \cite{Hergert:2012nb}. Because all irreducible many-body density matrices vanish in closed-shell nuclei, our findings indicate that the truncation of terms containing $\lambda^{(n\geq3)}$ and non-linear powers of $\lambda^{(2)}$ is negligible compared to the truncation of induced three-body operators. A more detailed analysis of the MR-IM-SRG truncation scheme will be presented in a future publication. 

%%%%%%%%%%%%%%%%%%%%%%%%%%%%%%%%%%%%%%%%%%%%%%%%%%%%%%%%%%%%%%%%%%%%%%
%%%%%%%%%%%%%%%%%%%%%%%%%%%%%%%%%%%%%%%%%%%%%%%%%%%%%%%%%%%%%%%%%%%%%%
%%%%%%%%%%%%%%%%%%%%%%%%%%%%%%%%%%%%%%%%%%%%%%%%%%%%%%%%%%%%%%%%%%%%%%
% SECTION %%%%%%%%%%%%%%%%%%%%%%%%%%%%%%%%%%%%%%%%%%%%%%%%%%%%%%%%%%%%
%%%%%%%%%%%%%%%%%%%%%%%%%%%%%%%%%%%%%%%%%%%%%%%%%%%%%%%%%%%%%%%%%%%%%%
\paragraph{Conclusions.}
We have generalized the IM-SRG approach to multi-reference states, and used the resulting MR-IM-SRG method to perform the first \emph{ab initio} study of all even oxygen isotopes with chiral NN+3N Hamiltonians, along with the IT-NCSM and the CC method. The MR-IM-SRG results are in excellent agreement with those from the other methods, confirming its reliability, and the method's modest computational demands make it ideally suited for the description of medium- and heavy-mass open-shell nuclei far from shell closures. 

Our calculated oxygen ground-state energies agree remarkably well with experimental binding energies within theoretical uncertainties of 3\%. This is achieved without any re-adjustment of the interaction to experimental data beyond $\nuc{He}{4}$, and therefore constitutes an impressive demonstration of the predictive power of chiral NN+3N Hamiltonians. The present work also highlights the importance of the 3N interaction for the nuclear structure of neutron-rich nuclei, as demonstrated by the robust reproduction of the oxygen drip line. 

%%%%%%%%%%%%%%%%%%%%%%%%%%%%%%%%%%%%%%%%%%%%%%%%%%%%%%%%%%%%%%%%%%%%%%
%%%%%%%%%%%%%%%%%%%%%%%%%%%%%%%%%%%%%%%%%%%%%%%%%%%%%%%%%%%%%%%%%%%%%%
%%%%%%%%%%%%%%%%%%%%%%%%%%%%%%%%%%%%%%%%%%%%%%%%%%%%%%%%%%%%%%%%%%%%%%
\paragraph{Acknowledgments.}
We thank S. Bogner and R. Furnstahl for useful discussions and comments. This work is supported in part by the National Science Foundation under Grant No.~PHY-1002478, and the NUCLEI SciDAC Collaboration under the U.S. Department of Energy Grant No. DE-SC0008533, the Deutsche Forschungsgemeinschaft through contract SFB 634, the Helmholtz International Center for FAIR (HIC for FAIR), and the BMBF through contract 06DA7074I. Computing resources were provided by the Ohio Supercomputer Center (OSC), the J\"ulich Supercomputing Center, the LOEWE-CSC Frankfurt, and the National Energy Research Scientific Computing Center supported by the Office of Science of the U.S. Department of Energy under Contract No. DE-AC02-05CHH11231.

%%%%%%%%%%%%%%%%%%%%%%%%%%%%%%%%%%%%%%%%%%%%%%%%%%%%%%%%%%%%%%%%%%%%%%
%%%%%%%%%%%%%%%%%%%%%%%%%%%%%%%%%%%%%%%%%%%%%%%%%%%%%%%%%%%%%%%%%%%%%%
%%%%%%%%%%%%%%%%%%%%%%%%%%%%%%%%%%%%%%%%%%%%%%%%%%%%%%%%%%%%%%%%%%%%%%
% \bibliography{/Users/hergert/Library/texmf/bibtex/bib/master_bibdesk}

%%%%%%%%%%%%%%%%%%%%%%%%%%%%%%%%%%%%%%%%%%%%%%%%%%%%%%%%%%%%%%%%%%%%%%
%%%%%%%%%%%%%%%%%%%%%%%%%%%%%%%%%%%%%%%%%%%%%%%%%%%%%%%%%%%%%%%%%%%%%%
%%%%%%%%%%%%%%%%%%%%%%%%%%%%%%%%%%%%%%%%%%%%%%%%%%%%%%%%%%%%%%%%%%%%%%
\end{document}